\documentclass[twocolumn]{article}
\usepackage{epsfig}
\usepackage{amsfonts}

\title{Measurement of the $\eta$ Production in Proton Proton Collisions with
  the COSY Time of Flight Spectrometer}

\author{
  M.~Abdel-Bary$^5$,  S.~Abdel-Samad$^5$,  R.~Bilger$^8$,
  K.-Th.~Brinkmann$^3$,  H.~Clement$^8$,\\    S.~Dshemuchadse $^6$,
  E.~Dorochkevitch $^8$,  H.~Dutz$^2$,  W.~Eyrich$^4$,  A.~Erhardt$^8$, 
  D.~Filges$^5$,\\  A.~Filippi$^7$, H.~Freiesleben$^3$,  M.~Fritsch$^4$,
  R.~Geyer$^5$,  A.~Gillitzer$^5$,  D.~Hesselbarth$^5$,\\  B.~Jakob$^3$,
  L.~Karsch$^3$,  K.~Kilian$^5$,  H.~Koch$^1$,  J.~Kre ss{}$^8$,
  E.~Kuhlmann$^3$,  S.~Marcello$^7$,  \\ S.~Marwinski$^5$,  S.~Mauro$^1$,
  W.~Meyer$^1$,  P.~Michel$^6$,  K.~M\"oller$^6$,  H.~P.~Morsch$^5$, \\
  L.~Naumann$^6$,  N.~Paul$^5$,  M.~Richter$^3$, 
  E.~Roderburg$^5$ \footnote{email-address: e.roderburg@fz-juelich.de}, 
  M.~Rogge$^5$,  A.~Schamlott$^6$,\\   M.~Schmitz$^5$,  P.~Sch\"onmeier$^3$,
  M.~Schulte-Wissermann$^3$,   W.~Schroeder$^4$,  T.~Sefzick$^5$, \\
  F.~Stinzing$^4$,  G.Y.~Sun$^3$,  G.J.~Wagner$^8$,  M.~Wagner$^4$, 
  A.~Wilms$^1$,  P.~Wintz$^5$, \\ S.~Wirth$^4$,  P.~Zupranski$^9$ \\
{\small $^1$  Institut f\"ur Experimentalphysik, Universit\"at Bochum, D-44780 Bochum} \\
{\small $^2$ Physikalisches Institut, Universit\"at Bonn, D-53115 Bonn} \\
{\small $^3$  Institut f\"ur Kern- und Teilchenphysik, Technische Universit\"at Dresden, D-01062 Dresden} \\ 
{\small $^4$ Physikalisches Institut, Universit\"at  Erlangen-N\"urnberg, D-91058 Erlangen }\\ 
{\small $^5$ Institut f\"ur Kernphysik,  Forschungszentrum J\"ulich, D-52425 J\"ulich }\\ 
{\small $^6$ Institut f\"ur Kern-  und Hadronenphysik, Forschungszentrum Rossendorf, D-01314 Dresden} \\ 
{\small $^7$ INFN Torino, I-10125 Torino }\\ 
{\small $^8$ Physikalisches Institut, Universit\"at  T\"ubingen, D-72076 T\"ubingen} \\ 
{\small $^9$ Andrzej Soltan Institute for Nuclear  Studies, PL-00681 Warsaw} }

\begin{document}

\maketitle

\begin{abstract} The reaction pp $\rightarrow$ pp$\eta$ was measured at
  excess energies of 15 and 41~MeV at an external target of the J\"ulich
  Cooler Synchrotron COSY with the Time of Flight Spectro\-meter.  About 25000
  events were measured for the excess energy of 15~MeV and about 8000 for
  41~MeV. Both protons of the process pp$\eta$ were detected with an
  acceptance of nearly 100 \% and the $\eta$ was reconstructed by the missing
  mass technique. For both excess
  energies the angular distributions are found  to be nearly isotropic.  In the invariant mass distributions strong
  deviations from the pure phase space distributions are seen.  
\end {abstract}

\section{Introduction}
The production of $\eta$ mesons in nucleon-nucleon collisions has attracted a
considerable attention in the last years \cite{Bergdolt}..\cite{Smyrski}.  Due
to the relatively high mass of the $\eta$ meson its production requires a
large momentum transfer between the initial and final states. At threshold
this momentum transfer is given by $
q=\sqrt(m_Nm_{\eta}+\frac{1}{4}m_{\eta}^2)$, where $m_N$ and $m_{\eta}$ are
the nucleon and $\eta$ meson masses, and amounts to 770 MeV/c.  This large
momentum transfer translates into a short distance between the two interacting
nucleons ($\sim 0.26$ fm).  Therefore $\eta$ production permits the investigation of short range hadron dynamics
and provides a tool to probe the behavior of nucleon-nucleon interaction at
small distances.  The mechanism of $\eta$ production in nucleon-nucleon
collisions is  yet not well understood. As the $\eta$ meson couples strongly to
the $S_{11}$(1535 MeV), it has been assumed that this resonance is
predominantly involved in its production. The mechanism of the excitation of
this resonance in hadronic collisions remains in turn an open question. As the
dominant excitation mechanism both $\pi$ and $\eta$ exchange
have been assumed \cite{Botovic}.  According to other calculations however,
$\rho$ meson exchange is supposed to be responsible mainly for the excitation
of the $S_{11}$ resonance \cite{Bernard}, \cite{Faldt}.  The
shape of the angular distribution in the $pp \rightarrow pp\eta$ reaction 
reflects the properties of the exchanged meson in the excitation of the
$S_{11}$ resonance. In Ref. \cite{Nakayama} it is shown that in spite of the
dominance of the resonance amplitude the effects of relatively small mesonic
currents can be seen in the shape of the $\eta$ angular distribution due to
interference. On the other hand, according to a recent approach \cite{Pena}
the $S_{11}$ resonance is less involved in the production of the $\eta$ meson.
In this approach the $\eta$ meson production is mainly due to the short range
amplitude of the nucleon-nucleon interaction. As the $\eta$ is the lightest meson
containing $s\overline{s}$ quarks the question of direct production channels
is interesting because it may be related to the $s\overline{s}$ content of
the proton.  The production of the $\eta$ meson also poses another interesting
aspect.  The $\eta$ meson interacts strongly with the nucleon and in its
production the final state $\eta$-nucleon interaction should play an important
role.  The near threshold dependence of the observed total cross section for
$\eta$ production differs substantially from that for $\pi$ and $\eta'$
production, pointing to an enhancement of the measured cross section which can
be ascribed to a strong attraction in $\eta$-nucleon interaction \cite{pawelCosy11}.  

More data of improved precision will undoubtedly contribute to a better understanding of
the $pp \rightarrow pp\eta$  reaction.
\section{Experiment}
\begin{figure}[htb]
\begin{center}
  \epsfig{file=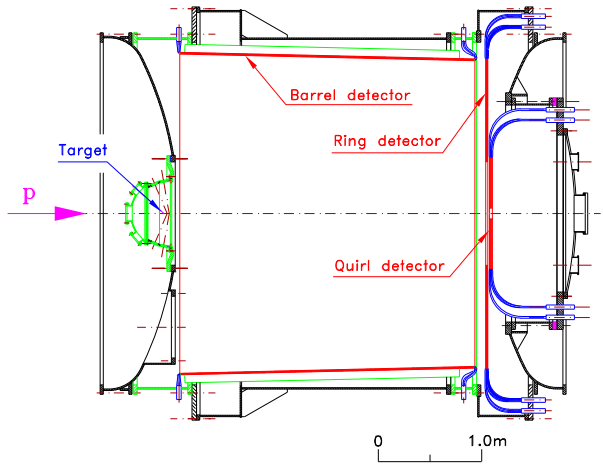,height=6cm}
     \caption{\label{experimentalsetup} The experimental set up of the COSY-TOF detector. 
       A cross section through the cylindrical detector is shown.}
\end{center}
\end{figure}

The external proton beam with a diameter of 1-2~mm impinges on a 4~mm thick
liquid hydrogen target enclosed with plastic foils of only 0.9~$\mu m$
thickness \cite{Target}.  The start and stop counters have azimuthal symmetry
with holes for the beam such that only reaction particles hit the counters.
Three scintillators with holes of 8~mm, 5~mm and 2~mm diameter placed 1~m,
0.5~m and 10~cm in front of the target are used to veto the beam halo.  The
incident beam and the whole detection system are enclosed in vacuum in order
to minimize multiple scattering and secondary reactions (Fig.
\ref{experimentalsetup}).
A 0.5 mm thick scintillator counter is used as start counter. It has an inner
hole of 3 mm diameter and is arranged in two cones each of 16 individual
counters \cite{Startzaehler}.  The stop counters are divided into the inner
part (quirl counter with 96 channels in 3 layers\cite{Quirl}), the ring
counter with 192 channels in 3 layers and the barrel counter with 192 channels
\cite{Barrel}. The high granularity of the stop scintillators results in a
resolution for the azimuthal angle of $2^\circ$ - $4^\circ$ and for the
polar angle of typically $0.5^\circ$. The acceptance for charged
particles in $\phi$ is $2\pi$, for $\theta$ it ranges from $0.8^\circ$ -
$75^\circ$. The time of flight is measured with a resolution of 1\% of the
typical 
time of flight for protons of the $pp
\rightarrow pp\eta$ reaction close to threshold.  For the measurement with a
beam momentum of 2025 MeV/c (corresponding to an excess energy $\epsilon$ of
the pp$\eta$ system of 15~MeV) the trigger condition was to measure at least
two charged tracks in the quirl and at least one charged track in the start
counter. For the beam momentum of 2100 MeV/c ($\epsilon$ = 41~MeV) the charged
tracks in the ring counter were added to the trigger in order to take into
account the larger opening angle of the protons.
\section{Data Analysis}
The $\eta$ is reconstructed from the tracks of both protons.  For each track
the particle momentum is calculated from the time of flight and the measured
directions with the mass of the proton assumed. The missing mass is calculated
from momentum and energy conservation as
$$
m_x^2 =(\mathbb{P}_{beam} + \mathbb{P}_{target} -\mathbb{P}_{track1}
-\mathbb{P}_{track2})^2
$$
$ \mathbb{P}$ is the four momentum of the particles.  Only two charged
track events are taken into account, such that events with the $\eta$ decaying
into charged particles hitting the ring or quirl counter are discarded.  Pion
tracks result in very low or negative squared missing masses if the mass of
the proton is assumed.  Near the kinematical limit, which corresponds to a
missing mass of 562.5 MeV/$c^2$ ($\epsilon$ =15~MeV) and 588.5 MeV/$c^2$
($\epsilon$ =41~MeV), only proton tracks from pp$\eta$ and multi-pion
production events are contributing. In order
to reduce the amount of background events, a first selection on the squared missing
mass is applied ($m_x^2 > 0.2$~$GeV^2/c^4$). This cut and the condition on the
particle time of flight corresponding to velocities of $0<\beta<1$ are the
only cuts applied in the analysis.

\begin{figure}[htb]
\begin{center}
  \epsfig{file= 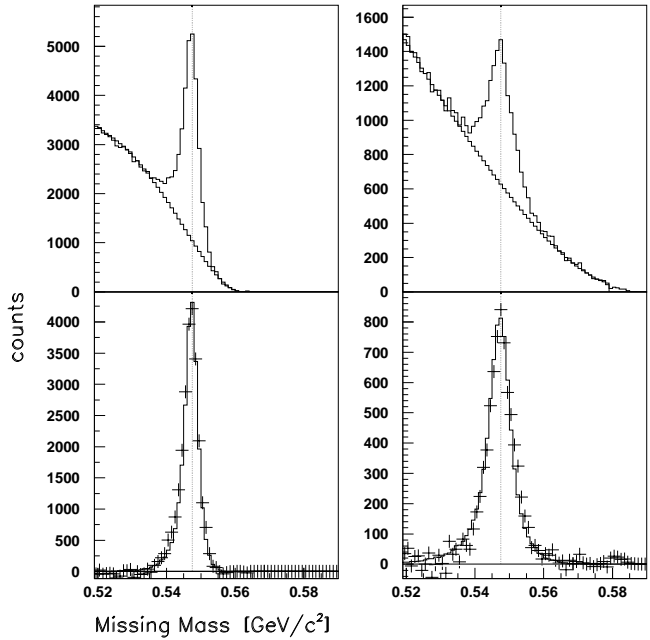,width=9. cm}
\caption{\label{missing-mass-data-mc} The missing mass distribution
  for the excess energy of 15~MeV (first column) and 41~MeV (second column). Upper
  figure: Raw data with fitted background.  Lower figure: Missing mass
  distribution from Monte Carlo calculations (solid line) in comparison with
  data after background subtraction (crosses).}
\end{center}
\end{figure}

The measured data are compared with simulated data obtained with a Monte Carlo
program, which is based on the GEANT3 code. It generates a data sample which
is analyzed using the same evaluation routines as for the measured data. As
 time resolution for each layer of the scintillation counters  250
ps ($\sigma$) for the stop counters and 150 ps ($\sigma$) for the start
counters was determined. The width of the $\eta$ signal in the missing mass distribution is
 well reproduced for both excess energies  (Fig. \ref{missing-mass-data-mc}).
The width of the $\eta$ signal is 0.8\% (FWHM) of the $\eta$ mass for
$\epsilon$=15~MeV and 1.3\% for $\epsilon$=41~MeV. The overall reconstruction
efficiency including the acceptance and the constraint of having no charged
pions of the $\eta$ decay in the forward detector was 68\% for $\epsilon$=15
MeV and 69\% for $\epsilon$=41~MeV.  
The distribution of the invariant proton-proton mass is para\-meterized by the
measured distribution and used instead of the pure phase space  as
input for the Monte Carlo calculation.
This influences the Monte Carlo results mainly at
forward - backward $\eta$-directions for the lower excess energy. 

\begin{figure}[htb]
\begin{center}
  \epsfig{file=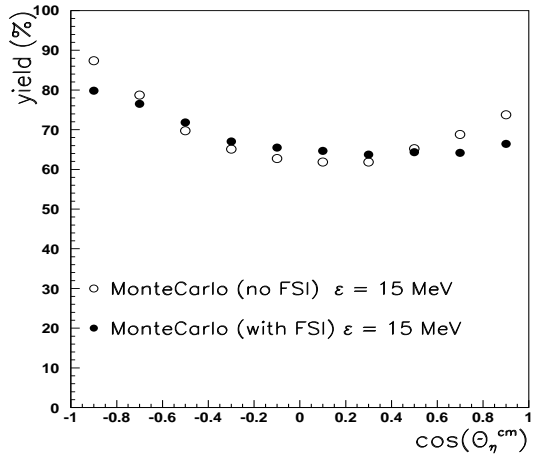,height=6cm,width=7.0 cm}
     \caption{\label{wineta-mc} Monte Carlo results: Acceptance and
       reconstruction efficiency as a function of $cos(\theta_{\eta}^{cm})$.}
\end{center}   
\end{figure}

The normalization
of the angular distributions affected by this correction amounts to less than
10 \% (s. Fig.  \ref{wineta-mc}). For the excess energy of 41~MeV this
correction is less than 2\%.

The pp$\eta$ system in the final state is described by 12 variables,  of
which 7 are fixed by momentum and energy conservation and the particle masses.
2 variables are given by the invariant masses $M_{pp}$
and $M_{p\eta}$. The remaining 3 variables describe the orientation of the
pp$\eta$ system with respect to the incoming proton axis. For these variables we
choose the angle of the $\eta$  momentum in the center of mass frame with respect to the beam axis
($\theta_{\eta}^{cm}$) and the angle of the relative momentum of the proton
proton system to the beam axis ($\theta_q^{cm}$).  The remaining variable is the
azimuthal orientation, which is of no significance  in this unpolarized measurement.

\begin{figure*}[hbt]
\begin{center}
  \epsfig{file=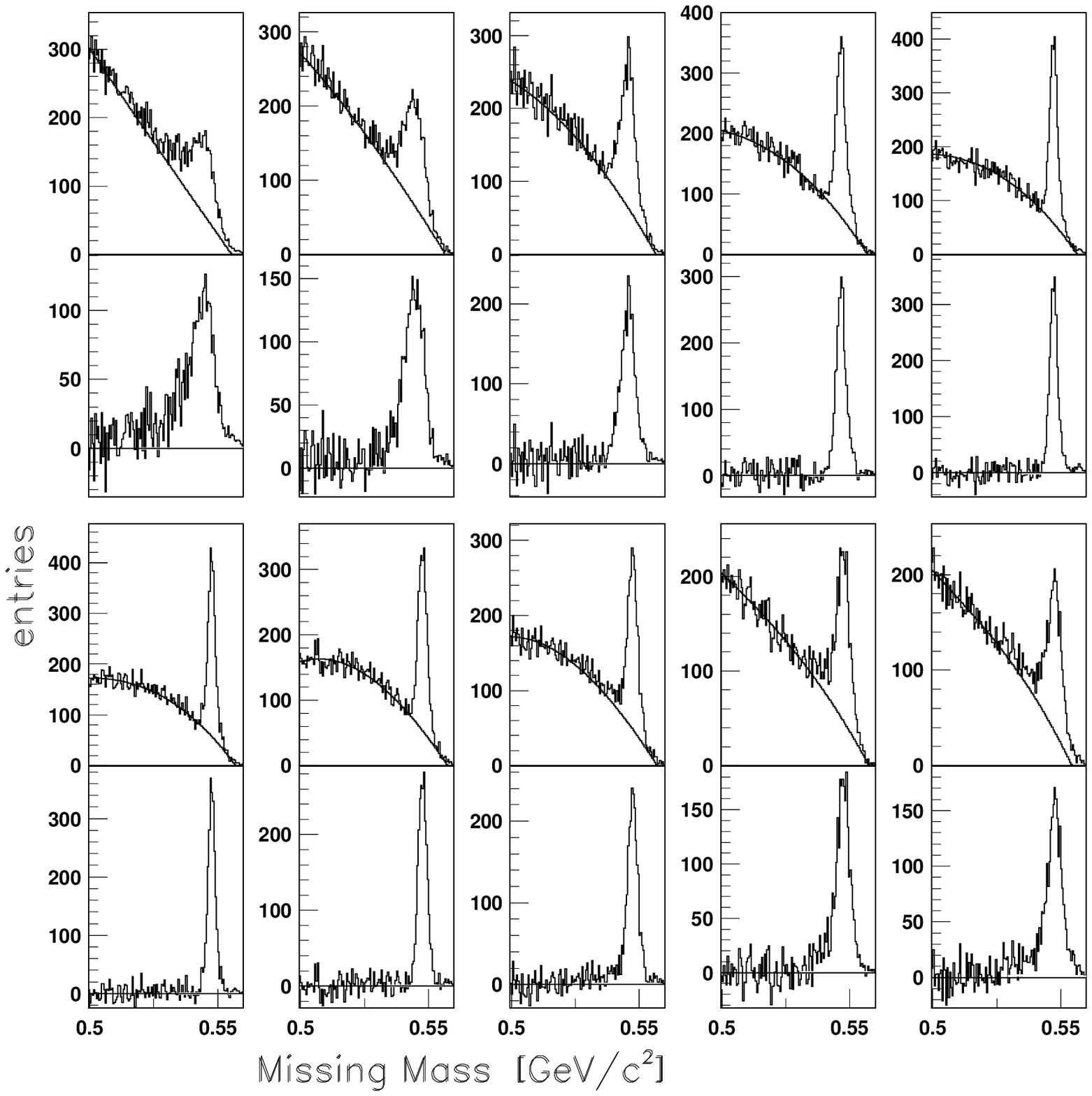 , width=15cm,height= 7. cm}
  \epsfig{file=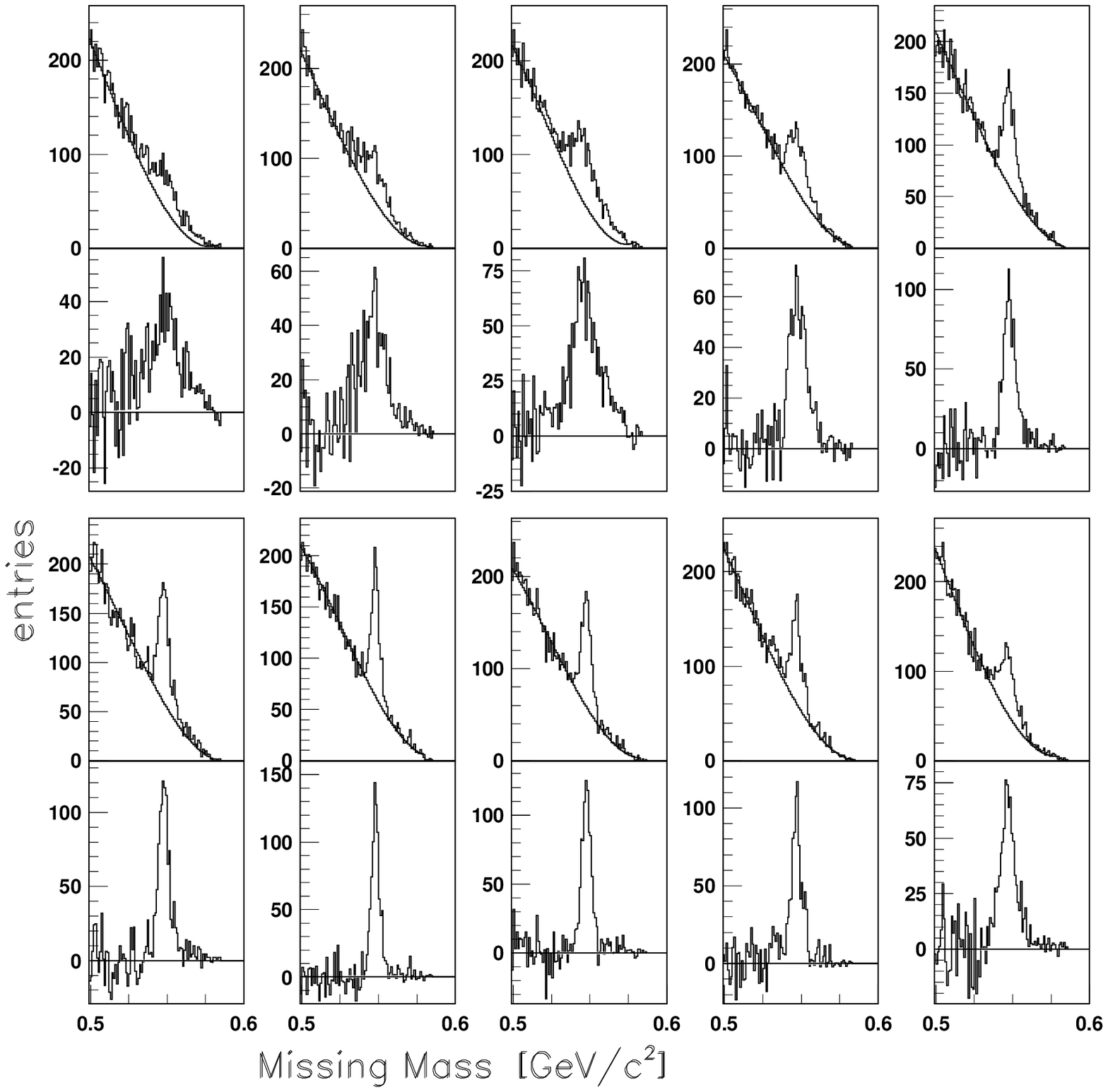 , width=15cm,height= 7. cm}
     \caption{\label{winetaueberblick} The missing mass
       distribution for bins of $cos(\theta_{\eta}^{cm})$ ($\epsilon$=15~MeV,   
       upper part and $\epsilon$ =41~MeV, lower part). The first row             
       shows the data with the fit of the background for                          
       $cos(\theta_{\eta}^{cm})$ in the range from -1. to 0. in steps of .2, the second   
       row shows the $\eta$ signal which is obtained by subtracting the fitted    
       background. The third and fourth rows show the corresponding bins of       
       $cos(\theta_{\eta}^{cm})$ in the range from 0. to 1.}                                      
\end{center} 
\end{figure*}

In order to obtain differential cross sections with respect to
the quantities $M^2_{pp}$, $M^2_{p\eta}$, $cos(\theta_{\eta}^{cm})$, and
$cos(\theta_q^{cm})$ for $pp \rightarrow pp \eta$ the continuum background in
the pp missing mass has to be subtracted. The procedure is illustrated in the
appendix, in case of $cos(\theta_{\eta}^{cm})$ in Fig.
\ref{winetaueberblick}. For each of 10
equidistant bins between $cos(\theta_{\eta}^{cm})=-1$ and
$cos(\theta_{\eta}^{cm})=+1$ pp missing mass spectra were generated. In each
missing mass distribution the number of $\eta$ events is determined by fitting
a polynomial of 3rd degree to the background and a Gaussian distribution to
the $\eta$ signal. The fit is done recursively by removing the fitted signal
from the overall measured distribution and fitting the background again.  The
width of the $\eta$ signal in the missing mass varies with the center of mass
angle of the $\eta$ due to the corresponding different velocities of the
protons resulting in different missing mass resolutions.  Therefore the limits
for integrating the $\eta$ signal are varying. These
limits are determined from the width of the $\eta$ signal of Monte Carlo
calculations.  

In the same way differential cross section distributions for
$M^2_{pp}$, $M^2_{p\eta}$, and $cos(\theta_q^{cm})$ were obtained by fitting
background and $\eta$ peak in the pp missing mass and subtracting the
background as a function of $M^2_{pp}$ ,$M^2_{p\eta}$, and $cos(\theta_q^{cm})$
respectively.

  The angular distributions are normalized to the total cross
sections of 2.11 $\mu b$ ($\epsilon$=15~MeV) and 4.92 $\mu b$
($\epsilon$=41~MeV)
 which are taken from Ref. \cite{Calen}.

In Figs. \ref{win-back}-\ref{inv-norm} only statistical errors are plotted. The systematic errors
are given in the appendix in tables \ref{tab-win}-\ref{tab-petainv}. The
estimated systematic error is composed of a constant term due to
possible angular dependent detector and reconstruction  inefficiencies ($\pm$5\%) and of a term which
is proportional to the width of the $\eta$ signal in the missing mass, in order to reflect
the error due to the separation of the signal from background ($\pm$2\%$\cdot$FWHM[MeV/$c^2$]).
\section{Angular Distributions}
\begin{figure}[htb]
\begin{center}
  \epsfig{file=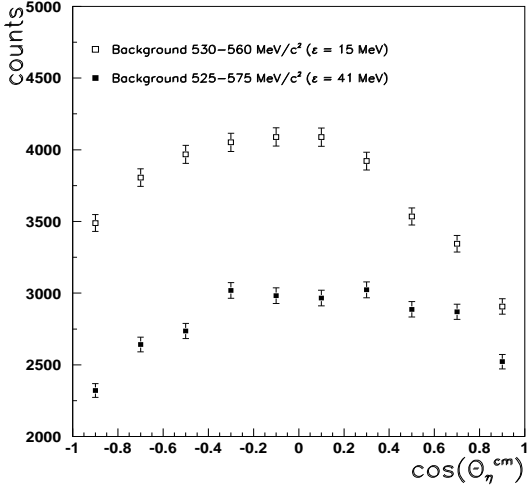,height=6.5cm,width=7.0 cm}
     \caption{\label{win-back} Number of background events as a function of  
       the cosine of the $\eta$ angle.}
\end{center} 
\end{figure}

The  angular distribution of the $\eta$ in the center of mass  system is evaluated as
described above.  This
procedure allows in addition for the examination of the background behavior,
which is shown in Fig.~\ref{win-back}.
The angular distribution of the background, which is measured for a constant range in the missing mass
spectrum, deviates from isotropy. 

\begin{figure}[htb]
\begin{center}
  \epsfig{file=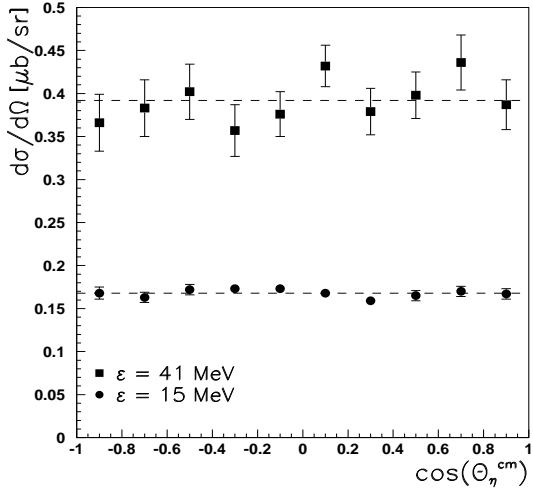,height=6.5cm,width=7.0 cm}
     \caption{\label{wineta} Angular distribution
       of the $\eta$ in the center of mass frame. }
\end{center} 
\end{figure}

The efficiency and acceptance corrected $\eta$ angular distributions for $\epsilon$=15~MeV and 41~MeV are shown
in Fig. \ref{wineta}.
Within the statistical errors both distributions show no deviation from
isotropy.  
The forward-backward symmetry in the center of mass frame required by the
initial state consisting of two identical particles  has not been used as a constraint in
the data evaluation.

The second variable describing the orientation of the pp$\eta$ system is the
angle $\theta_q^{cm}$.
The relative momentum between both protons in the overall center of mass frame is given
by:
$$\vec q=\frac{1}{2}\cdot(\vec p\,^{cm}_{proton_1}-\vec p\,^{cm}_{proton_2})$$
 
\begin{figure}[htb]
\begin{center}
  \epsfig{file=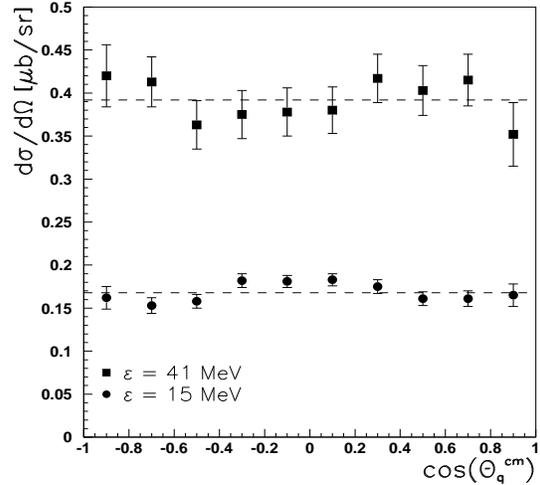,height=6.5cm,width=7.0 cm}
     \caption{\label{winq} Angular distribution
       of the relative p p momentum in the center of mass frame. }
\end{center} 
\end{figure}

The sequence of first and second proton is randomized for evaluating the angular distribution which is
shown in Fig. \ref{winq}.

As in the case of the $\eta$ angular distribution these differential cross
sections  are consistent with isotropy.

\section{Invariant Mass Distributions}
The invariant mass distributions are evaluated as explained in
chapter 4. The sequence of missing mass spectra is shown in Figs. 
\ref{ppinv15ueberblick},\ref{ppinv41ueberblick} in the appendix.
The squared proton-proton invariant mass is given by

$$
M_{pp}^2 =(\mathbb{P}_{track_1} +\mathbb{P}_{track_2})^2
$$

\begin{figure}[htb]
\begin{center}
  \epsfig{file=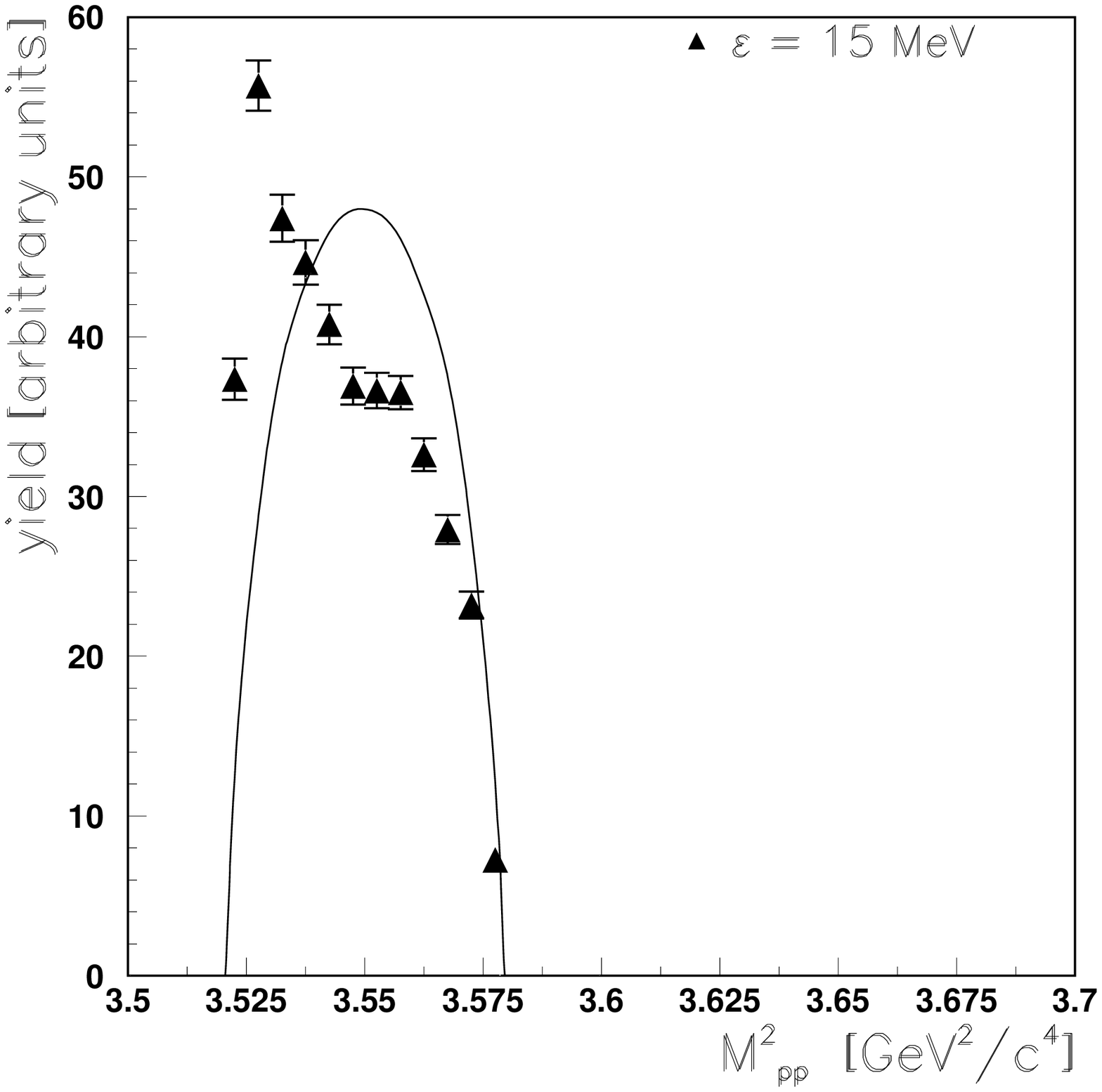,width=7.0cm,height=6.50cm}
  \epsfig{file=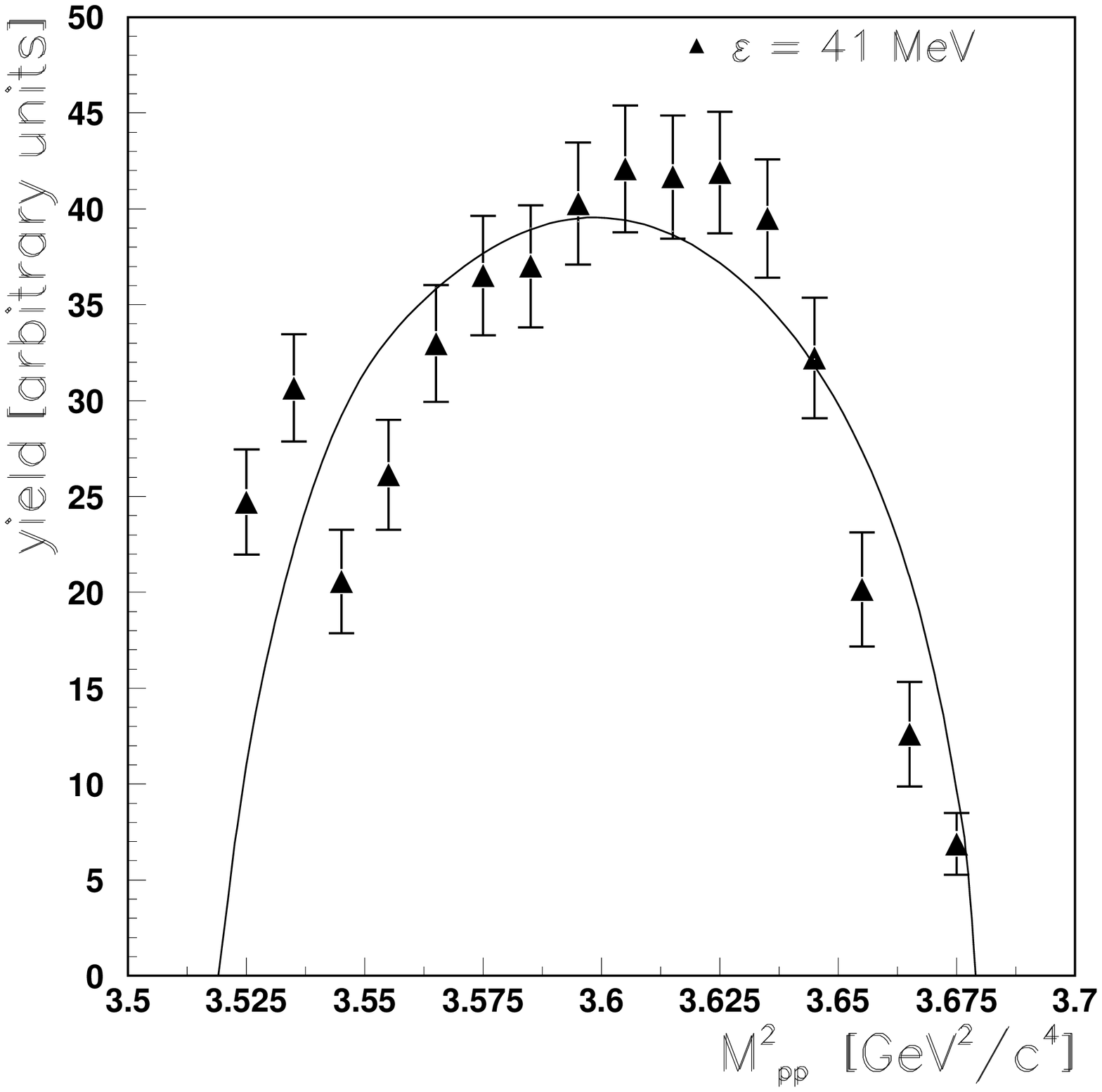,width=7.0cm,height=6.50cm}
     \caption{\label{ppinv+lines} Squared proton-proton 
       invariant mass distribution (upper Figure: excess energy of 15~MeV,
       lower Figure: excess energy of 41~MeV). The solid points show the
       efficiency and acceptance corrected data.
       The line represents the pure phase space.}
\end{center}
\end{figure}

The efficiency and acceptance corrected  proton-proton invariant mass
distributions are shown in Fig.\ref{ppinv+lines}. 

The invariant mass of the proton - $\eta$ system is given by
$$
M_{p\eta}^2=(\mathbb{P}_{\eta} +\mathbb{P}_{track_{1,2}})^2
$$
The four momentum of the $\eta$ particle is calculated from the momentum
$$
\vec{p}_{\eta} = \vec{p}_{beam}-\vec{p}_{track_1}-\vec{p}_{track_2}$$
and
from the total energy
$$E = \sqrt{(p_{\eta}^2+m_{\eta}^2)}$$
For $ m_{\eta}$ the exact mass (.5475 GeV/$c^2$) is taken. As the protons of
the final state are not distinguishable the
invariant mass $M_{p\eta}^2$ is evaluated twice for each event by exchanging
track 1 with track 2.

\begin{figure}[htb]
\begin{center}
  \epsfig{file=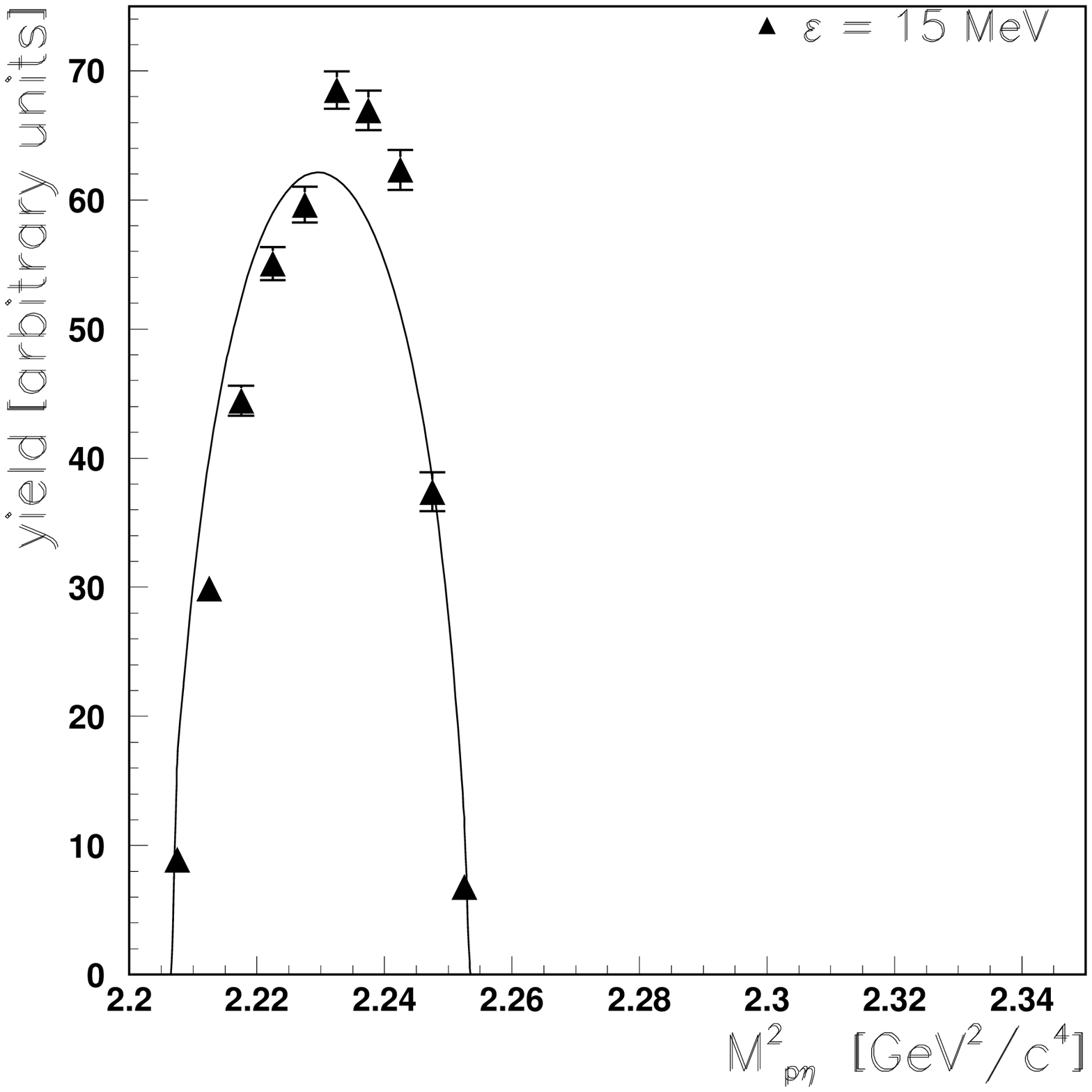,width=7.0cm,height=6.50cm}
  \epsfig{file=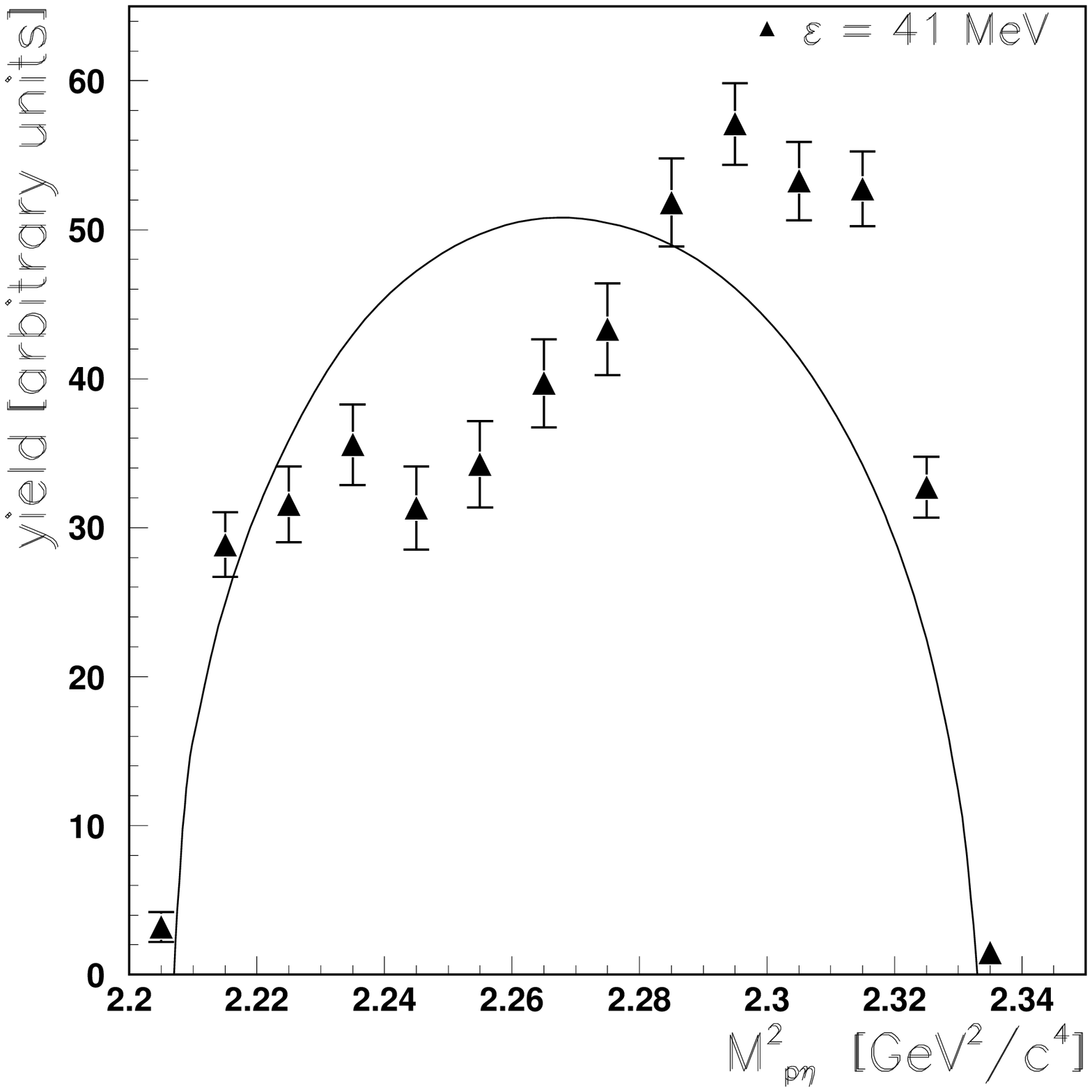,width=7.0cm,height=6.50cm}
     \caption{\label{petainv+lines} Squared proton-$\eta$ 
       invariant mass distribution (upper Figure: excess energy of 15~MeV,
       lower Figure: excess energy of 41~MeV). The solid points show the  efficiency and acceptance corrected  data.
       The line represents 
       the  pure phase space. }
\end{center} 
\end{figure}

The proton-$\eta$ invariant mass distributions, which  are deduced in the same
 way as the proton-proton invariant mass distributions, are shown  
 in Fig. \ref{petainv+lines}.

The numerical values are listed in the appendix in  tables \ref{tab-ppinv}, \ref{tab-petainv}. The systematic errors
are calculated as explained in section 3.

\section{Discussion}

The angular distributions at the excess energy of $\epsilon$=41~MeV were
fitted with a function containing a $cos^2\theta$
dependence: $c_1+c_2\cdot cos^2\theta$. For the $cos\theta_{\eta}$
distribution the results were $c_1=0.396 \pm 0.009$ and $c_2=-0.011 \pm 0.032$, for the
$cos\theta_q$ distribution the results were $c_1=0.386 \pm 0.014$ and $c_2=0.018 \pm
0.035$. 
 This means that the measured angular distributions are isotropic and
therefore consistent with s-wave behavior.

\begin{figure}[htb]
\begin{center}
  \epsfig{file=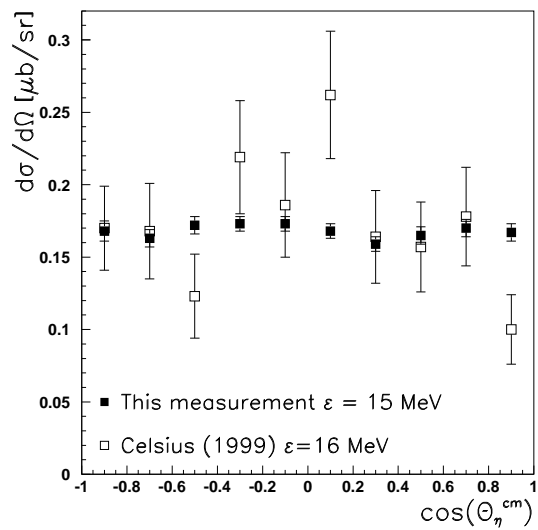 ,height=7.cm, width=7.0cm }
 
     \caption{\label{winetacompare15}Comparison of the measured angular distribution
with the measurement of CELSIUS \cite{Calen3}  close to $\epsilon$=15~MeV}
\end{center} 
\end{figure}

\begin{figure}[htb] 
\begin{center}
  \epsfig{file=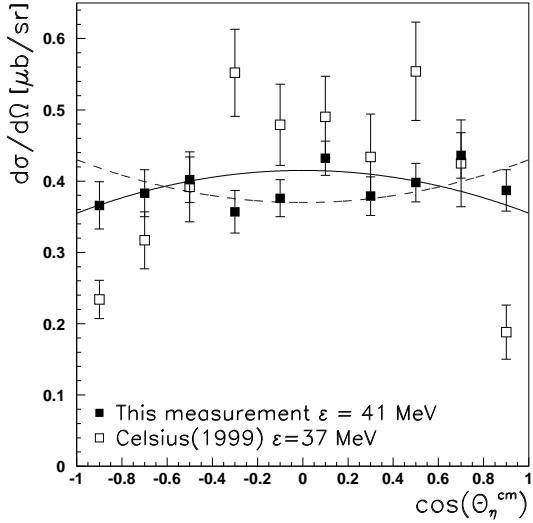 ,height=7.cm,width=7.0cm }
 
     \caption{\label{winetacompare41}Comparison of the measured angular distribution
with the measurement of CELSIUS \cite{Calen3} close to $\epsilon$=40 MeV.
The solid line represents calculations from Refs. \cite{Faldt}, \cite{Nakayama}, 
the dashed line shows the calculation from Ref. \cite{Nakayama} by only taking the
resonance excitation into account.}
\end{center} 
\end{figure}

This is in contradiction to the results from CELSIUS
\cite{Calen3} which are based on a data sample of lower statistical significance:
a factor of 80 (10) smaller for the lower (higher) excess energy. 
The data are compared with our results in Figs.
\ref{winetacompare15}, \ref{winetacompare41}. At the higher excess energy
close to $\epsilon$=40 MeV the shapes of the angular distributions are in
disagreement, particularly at the most forward and backward angles. 

Theoretical predictions for the angular distribution of G. F\"aldt, C. Wilkin
\cite{Faldt} and K. Nakayama \cite{Nakayama} are shown in Fig.
\ref{winetacompare41}.  In both calculations the excitation of the $S_{11}$
resonance is the dominant term. In the model of Ref. \cite{Nakayama} the $\pi$- and
$\eta$ exchange is the largest contribution of the resonance excitation, while
in Ref. \cite{Faldt}  $\rho$ exchange is dominating. While the results of both
calculations show nearly the same shape (solid line in Fig.
\ref{winetacompare41}), the origin of this shape is different. In Ref.
\cite{Nakayama} it arises from an interference between the very small non
resonant meson exchange process with the resonance excitation process, which -
if taken alone - produces an inverted shape as shown by the dashed line in
Fig. \ref{winetacompare41}. In Ref. \cite{Faldt} it is argued  that this shape is
due to the $\rho$ exchange of the resonance excitation and that a dominant
$\pi$ exchange will invert the shape.

Within the statistical and systematical error neither of the calculations can be excluded by the data. 

\begin{figure}[htb]
\begin{center}
  \epsfig{file=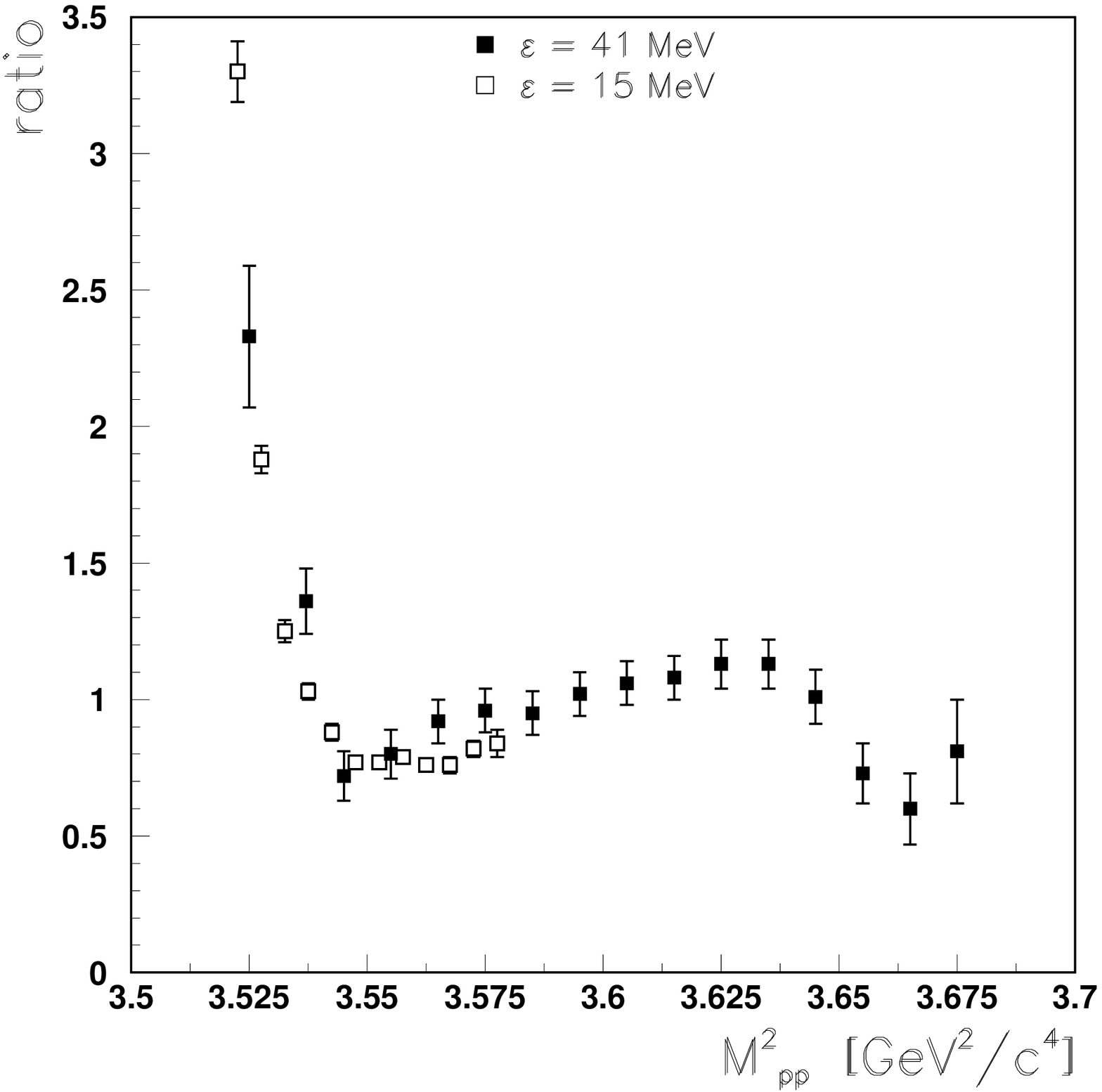 ,height=7.cm,  width=7.0cm }
 \epsfig{file=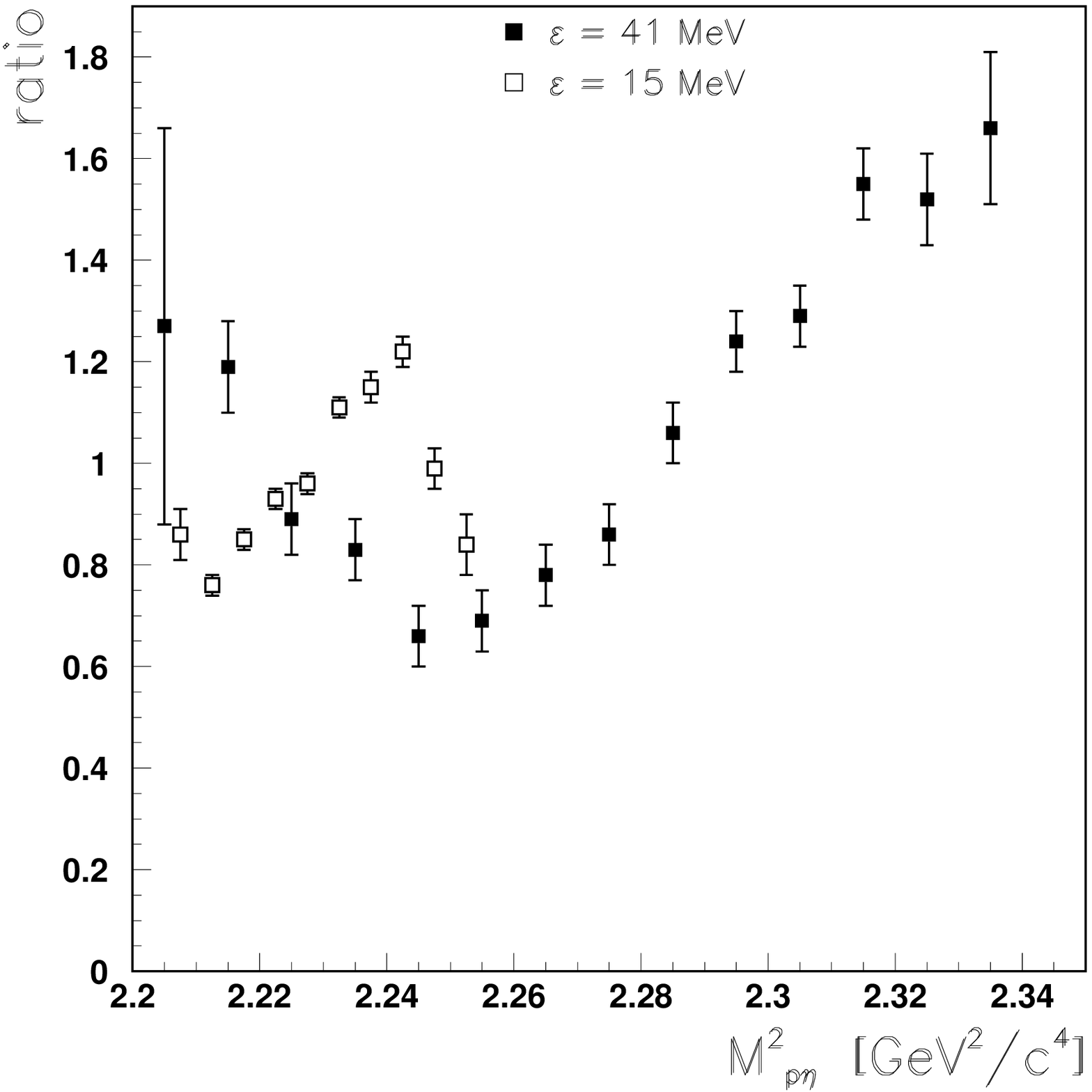 ,height=7.cm, width=7.0cm}
     \caption{\label{inv-norm} Ratio of the acceptance corrected measured
 invariant mass distribution and the pure phase space. Upper figure: proton-proton invariant 
mass, lower figure: proton-$\eta$ invariant mass  }
\end{center} 
\end{figure}

The invariant mass distributions exhibit large deviations from the pure phase
space as shown in Figs.  \ref{ppinv+lines},\ref{petainv+lines}, and in Fig.
\ref{inv-norm}, where the data are normalized to the phase space.  The first
maximum in the proton-proton invariant mass spectrum is due to the pp final
state interaction.  The effect, that a second maximum is seen for $\epsilon$ =
41 MeV is a surprising result and cannot be explained by an interplay of pp
FSI and phase space distribution alone. In the proton-$\eta$ distribution a
clear shift to higher energies compared to phase space is observed. An
enhancement towards higher invariant masses is expected either by a Breit -
Wigner parameterization for the $S_{11}(1535)$ or by a FSI approach with a
positive scattering length for the proton - $\eta$. But as simple weight
factors they cannot describe the strong modulations in the invariant proton -
$\eta$ mass spectrum. The Dalitz plot occupation probably has to be
parameterized by amplitude contributions in the 3 two body subsystems
taking care of interference effects.

\section{Conclusion}
The high statistics and high acceptance measurements of the $pp\rightarrow
pp\eta$ reaction were performed with the COSY-TOF spectrometer.  Kinematically
complete events were obtained. They allow for detailed analysis of all
possible observables. The differential cross sections of $\eta$ with respect
to the beam direction show that up to 40 MeV excess energy the $\eta$ is still produced
purely in s-wave. There are deviations of the invariant mass distributions
from the pure phase, which cannot be explained by pp FSI alone.

\section{Acknowledgment}
We thank the COSY Crew for delivering a stable and high quality proton beam.
 This work has been supported by the FFE fond of the
Forschungszentrum-J\"ulich, by the European Community - Access to Research
Infrastructure action of the Improving Human Potential Programme, and by
the German BMBF. One of us (P.Z.) wishes to thank Polish State Committee
for Scientific Research for a partial support under grant KBN 3P 03B 04521.

\onecolumn
\section{Appendix}

\begin{center}
\begin{figure*}
  \epsfig{file= 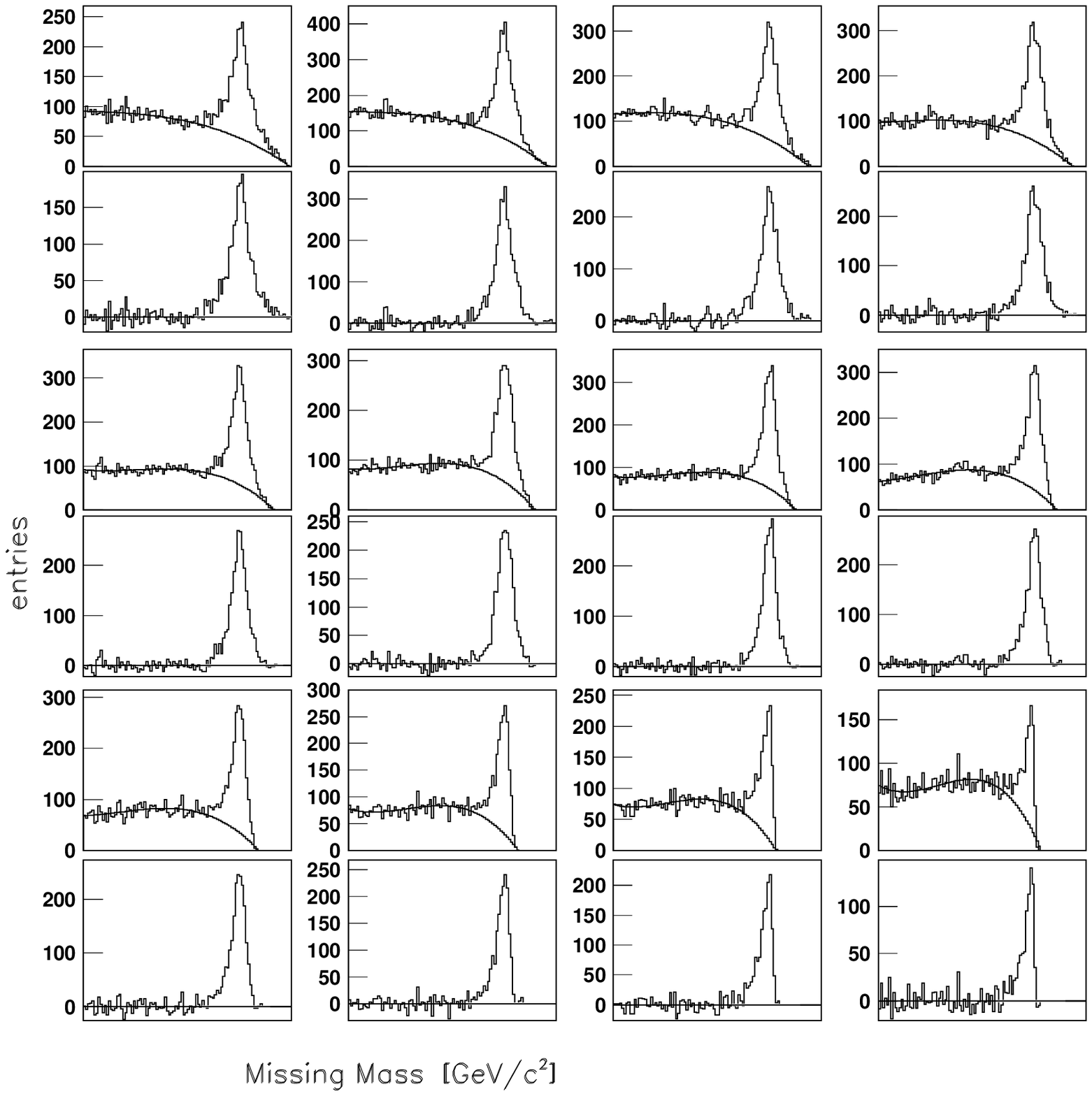,width=17cm,height= 17. cm}
 
     \caption{\label{ppinv15ueberblick} The missing mass 
       distribution for bins of the proton-proton invariant mass at
       $\epsilon=15$ MeV. The upper histogram shows the data including the fit
       of the background.  Beneath this histogram the $\eta$ signal is shown,
       which is obtained by subtracting the fitted background. Each histogram
       represents a bin in $M^2_{pp}$ in the range from 3.52 to 3.57 $GeV^2/c^4$ in steps of
       .005 $GeV^2/c^4$.}
\end{figure*}
\end{center} 
\begin{center}
\begin{figure*}
  \epsfig{file= 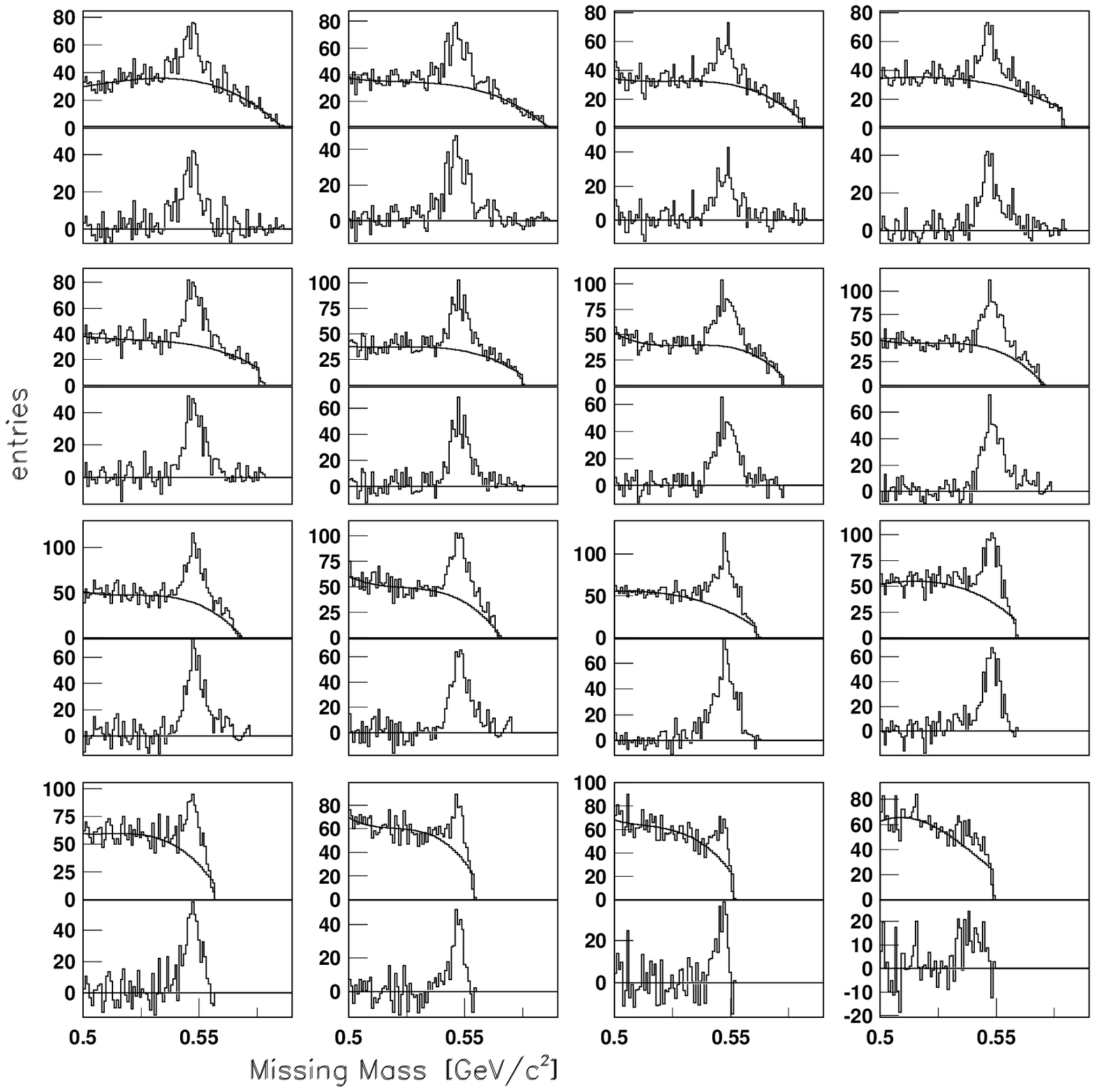,width=17cm,height= 16. cm}
 
     \caption{\label{ppinv41ueberblick} The missing mass 
       distribution for bins of the proton-proton invariant mass at
       $\epsilon=41$ MeV. The upper histogram shows the data including the fit
       of the background.  Beneath this histogram the $\eta$ signal is shown,
       which is obtained by subtracting the fitted background. Each histogram
       represents a bin in $M^2_{pp}$ in the range from 3.52 to 3.68 $GeV^2/c^4$ in steps of
       .01 $GeV^2/c^4$.}
\end{figure*}
\end{center} 


\begin{table}[htb]
\begin{center}
\caption {Angular distribution of the $\eta$ momentum in the cm system}
\label{tab-win}
\begin{tabular} {lllllll}
\hline\\

& \multicolumn{3}{c}{ \bf{$\epsilon $ = 15~MeV} }   &\multicolumn{3}{c}{ \bf{$\epsilon $ = 41~MeV} }   \\
\bf{$cos(\theta_{\eta}^{cm})$} &&&&&&\\
&$\frac{d\sigma}{d\Omega}$[$\mu$b/sr]& stat. error&  sys. error&$\frac{d\sigma}{d\Omega}$[$\mu$b/sr]& stat. error&  sys. error \\ 
&&&&&&\\
\hline \\
-0.9 &0.168 & $\pm$ 0.007& $\pm$0.018 &0.366 & $\pm$0.033 & $\pm$0.068 \\
-0.7 &0.163 & $\pm$ 0.006& $\pm$0.015 &0.383 & $\pm$0.033 & $\pm$0.064 \\
-0.5 &0.172 & $\pm$ 0.006& $\pm$0.013 &0.402 & $\pm$0.032 & $\pm$0.055 \\
-0.3 &0.173 & $\pm$ 0.005& $\pm$0.011 &0.357 & $\pm$0.030 & $\pm$0.040 \\
-0.1 &0.173 & $\pm$ 0.005& $\pm$0.011 &0.376 & $\pm$0.026 & $\pm$0.035 \\
+0.1 &0.168 & $\pm$ 0.005& $\pm$0.010 &0.432 & $\pm$0.024 & $\pm$0.036 \\
+0.3 &0.159 & $\pm$ 0.005& $\pm$0.009 &0.379 & $\pm$0.027 & $\pm$0.030 \\
+0.5 &0.165 & $\pm$ 0.006& $\pm$0.010 &0.398 & $\pm$0.027 & $\pm$0.034 \\
+0.7 &0.170 & $\pm$ 0.006& $\pm$0.012 &0.436 & $\pm$0.032 & $\pm$0.042 \\
+0.9 &0.167 & $\pm$ 0.006& $\pm$0.013 &0.387 & $\pm$0.029 & $\pm$0.045 \\
\hline\\
\end{tabular}

\caption {Angular distribution of the relative  momentum of the protons in the cm system}
\begin{tabular} {lllllll}
\hline \\
   & \multicolumn{3}{c}{ \bf{$\epsilon $ = 15~MeV} }   &\multicolumn{3}{c}{ \bf{$\epsilon $ = 41~MeV} }   \\
\bf{$cos(\theta_{q}^{cm})$} &&&&&&\\
&$\frac{d\sigma}{d\Omega}$[$\mu$b/sr]& stat. error&  sys. error&$\frac{d\sigma}{d\Omega}$[$\mu$b/sr]& stat. error&  sys. error \\ 
\hline \\
-0.9 &0.162 & $\pm$0.007& $\pm$0.014 &0.420 & $\pm$0.036 & $\pm$0.065 \\
-0.7 &0.153 & $\pm$0.006& $\pm$0.012 &0.413 & $\pm$0.029 & $\pm$0.056 \\
-0.5 &0.158 & $\pm$0.006& $\pm$0.012 &0.363 & $\pm$0.028 & $\pm$0.044 \\
-0.3 &0.182 & $\pm$0.006& $\pm$0.014 &0.375 & $\pm$0.028 & $\pm$0.041 \\
-0.1 &0.181 & $\pm$0.006& $\pm$0.013 &0.378 & $\pm$0.028 & $\pm$0.041 \\
+0.1 &0.183 & $\pm$0.006& $\pm$0.014 &0.380 & $\pm$0.027 & $\pm$0.041 \\
+0.3 &0.175 & $\pm$0.006& $\pm$0.013 &0.417 & $\pm$0.028 & $\pm$0.047 \\
+0.5 &0.161 & $\pm$0.006& $\pm$0.013 &0.403 & $\pm$0.029 & $\pm$0.049 \\
+0.7 &0.161 & $\pm$0.006& $\pm$0.013 &0.415 & $\pm$0.030 & $\pm$0.056 \\
+0.9 &0.165 & $\pm$0.007& $\pm$0.014 &0.352 & $\pm$0.037 & $\pm$0.053 \\
\hline\\
\end{tabular}
\end{center}
\end{table}
\vspace*{3cm}
\begin{table}[htb]
\begin{center}
\caption {Squared proton-proton invariant mass distribution }
\label{tab-ppinv}
\begin{tabular} {lllllll}
\hline
\multicolumn{7}{c}{\bf{$\epsilon$ = 15~MeV}}\\
& \multicolumn{3}{c}{ \bf{data} }   &\multicolumn{3}{c}{ \bf{data normalized to phase space} }   \\
$M_{pp}^2 [GeV^2/c^4]$&yield & stat. error& sys. error&
ratio& stat. error&sys. error\\
&[arb. units]&&&&&\\

\hline\\
  3.5225 & 37.33 & $\pm$1.29  &   $\pm$3.15 &3.30  &   $\pm$0.11 &    $\pm$ 0.28\\
  3.5275 & 55.71 & $\pm$1.58  &   $\pm$4.61 &1.88  &   $\pm$0.05 &    $\pm$ 0.16\\
  3.5325 & 47.41 & $\pm$1.47  &   $\pm$3.93 &1.25  &   $\pm$0.04 &    $\pm$ 0.10\\
  3.5375 & 44.65 & $\pm$1.38  &   $\pm$3.70 &1.03  &   $\pm$0.03 &    $\pm$ 0.09\\
  3.5425 & 40.78 & $\pm$1.24  &   $\pm$3.18 &0.88  &   $\pm$0.03 &    $\pm$ 0.07\\
  3.5475 & 36.91 & $\pm$1.16  &   $\pm$2.83 &0.77  &   $\pm$0.02 &    $\pm$ 0.06\\
  3.5525 & 36.63 & $\pm$1.10  &   $\pm$2.75 &0.77  &   $\pm$0.02 &    $\pm$ 0.06\\
  3.5575 & 36.50 & $\pm$1.05  &   $\pm$2.69 &0.79  &   $\pm$0.02 &    $\pm$ 0.06\\
  3.5625 & 32.61 & $\pm$1.02  &   $\pm$2.31 &0.76  &   $\pm$0.02 &    $\pm$ 0.05\\
  3.5675 & 27.93 & $\pm$0.92  &   $\pm$1.97 &0.76  &   $\pm$0.03 &    $\pm$ 0.05\\
  3.5725 & 23.20 & $\pm$0.85  &   $\pm$1.58 &0.82  &   $\pm$0.03 &    $\pm$ 0.06\\
  3.5775 &  7.26 & $\pm$0.43  &   $\pm$0.50 &0.84  &   $\pm$0.05 &    $\pm$ 0.06\\
        
\hline
\multicolumn{7}{c}{\bf{$\epsilon$ = 41~MeV}} \\
  3.525 & 24.71  &  $\pm$2.74 &    $\pm$3.53 &   2.33  &   $\pm$0.26  &    $\pm$0.33 \\     
  3.535 & 30.66  &  $\pm$2.80 &    $\pm$6.14 &   1.36  &   $\pm$0.12  &    $\pm$0.27 \\     
  3.545 & 20.56  &  $\pm$2.70 &    $\pm$3.02 &   0.72  &   $\pm$0.09  &    $\pm$0.11 \\     
  3.555 & 26.13  &  $\pm$2.87 &    $\pm$3.74 &   0.80  &   $\pm$0.09  &    $\pm$0.11 \\     
  3.565 & 32.97  &  $\pm$3.04 &    $\pm$4.53 &   0.92  &   $\pm$0.08  &    $\pm$0.13 \\     
  3.575 & 36.51  &  $\pm$3.11 &    $\pm$4.75 &   0.96  &   $\pm$0.08  &    $\pm$0.12 \\     
  3.585 & 37.01  &  $\pm$3.18 &    $\pm$4.61 &   0.95  &   $\pm$0.08  &    $\pm$0.12 \\     
  3.595 & 40.27  &  $\pm$3.18 &    $\pm$4.94 &   1.02  &   $\pm$0.08  &    $\pm$0.13 \\     
  3.605 & 42.09  &  $\pm$3.31 &    $\pm$4.93 &   1.06  &   $\pm$0.08  &    $\pm$0.12 \\     
  3.615 & 41.66  &  $\pm$3.21 &    $\pm$4.73 &   1.08  &   $\pm$0.08  &    $\pm$0.12 \\     
  3.625 & 41.89  &  $\pm$3.18 &    $\pm$4.68 &   1.13  &   $\pm$0.09  &    $\pm$0.13 \\     
  3.635 & 39.50  &  $\pm$3.09 &    $\pm$4.28 &   1.13  &   $\pm$0.09  &    $\pm$0.12 \\     
  3.645 & 32.22  &  $\pm$3.14 &    $\pm$3.26 &   1.01  &   $\pm$0.10  &    $\pm$0.10 \\     
  3.655 & 20.15  &  $\pm$2.97 &    $\pm$2.04 &   0.73  &   $\pm$0.11  &    $\pm$0.07 \\     
  3.665 & 12.59  &  $\pm$2.73 &    $\pm$1.23 &   0.60  &   $\pm$0.13  &    $\pm$0.06 \\                  
  3.675 &  6.88  &  $\pm$1.61 &    $\pm$0.72 &   0.81  &   $\pm$0.19  &    $\pm$0.08 \\                  
\hline\\                                             
\end{tabular}                                          
\end{center}
\end{table}
\vspace*{6cm}
\begin{table}[htb]
\begin{center}
\caption {Squared proton $\eta$ invariant mass distribution }
\label{tab-petainv}
\begin{tabular} {lllllll}
\hline
\multicolumn{7}{c}{\bf{$\epsilon$ = 15~MeV}}\\
& \multicolumn{3}{c}{ \bf{data} }   &\multicolumn{3}{c}{ \bf{data normalized to phase space} }   \\
$M_{p\eta}^2 [GeV^2/c^4]$&yield& stat. error& sys. error&ratio& stat. error&sys. error\\
&[arb. units]&&& &&\\

\hline
 2.2075 &  8.89  & $\pm$ 0.48 &   $\pm$ 0.60 &0.86 &  $\pm$0.05 &    $\pm$0.06 \\         
 2.2125 & 29.89  & $\pm$ 0.92 &   $\pm$ 2.07 &0.76 &  $\pm$0.02 &    $\pm$0.05 \\         
 2.2175 & 44.45  & $\pm$ 1.15 &   $\pm$ 3.14 &0.85 &  $\pm$0.02 &    $\pm$0.06 \\         
 2.2225 & 55.07  & $\pm$ 1.28 &   $\pm$ 3.97 &0.93 &  $\pm$0.02 &    $\pm$0.07 \\         
 2.2275 & 59.62  & $\pm$ 1.39 &   $\pm$ 4.30 &0.96 &  $\pm$0.02 &    $\pm$0.07 \\         
 2.2325 & 68.50  & $\pm$ 1.45 &   $\pm$ 4.84 &1.11 &  $\pm$0.02 &    $\pm$0.08 \\         
 2.2375 & 66.93  & $\pm$ 1.51 &   $\pm$ 4.64 &1.15 &  $\pm$0.03 &    $\pm$0.08 \\         
 2.2425 & 62.33  & $\pm$ 1.56 &   $\pm$ 4.23 &1.22 &  $\pm$0.03 &    $\pm$0.08 \\         
 2.2475 & 37.39  & $\pm$ 1.51 &   $\pm$ 2.54 &0.99 &  $\pm$0.04 &    $\pm$0.07 \\         
 2.2525 &  6.80  & $\pm$ 0.52 &   $\pm$ 0.47 &0.84 &  $\pm$0.06 &    $\pm$0.06 \\         
\hline

\multicolumn{7}{c}{\bf{$\epsilon$ = 41~MeV}} \\
 2.205 & 3.21 &  $\pm$1.00  &$\pm$0.34 &1.27 &   $\pm$0.39 &   $\pm$0.13 \\    
 2.215 &28.85 &  $\pm$2.17  &$\pm$3.02 &1.19 &   $\pm$0.09 &   $\pm$0.12 \\    
 2.225 &31.57 &  $\pm$2.55  &$\pm$3.53 &0.89 &   $\pm$0.07 &   $\pm$0.10 \\    
 2.235 &35.57 &  $\pm$2.71  &$\pm$3.98 &0.83 &   $\pm$0.06 &   $\pm$0.09 \\    
 2.245 &31.32 &  $\pm$2.78  &$\pm$3.61 &0.66 &   $\pm$0.06 &   $\pm$0.08 \\    
 2.255 &34.25 &  $\pm$2.91  &$\pm$4.01 &0.69 &   $\pm$0.06 &   $\pm$0.08 \\    
 2.265 &39.69 &  $\pm$2.96  &$\pm$4.80 &0.78 &   $\pm$0.06 &   $\pm$0.09 \\    
 2.275 &43.32 &  $\pm$3.07  &$\pm$5.16 &0.86 &   $\pm$0.06 &   $\pm$0.10 \\    
 2.285 &51.82 &  $\pm$2.95  &$\pm$5.98 &1.06 &   $\pm$0.06 &   $\pm$0.12 \\    
 2.295 &57.10 &  $\pm$2.75  &$\pm$6.28 &1.24 &   $\pm$0.06 &   $\pm$0.14 \\    
 2.305 &53.27 &  $\pm$2.63  &$\pm$5.77 &1.29 &   $\pm$0.06 &   $\pm$0.14 \\    
 2.315 &52.75 &  $\pm$2.51  &$\pm$5.52 &1.55 &   $\pm$0.07 &   $\pm$0.16 \\    
 2.325 &32.71 &  $\pm$2.03  &$\pm$3.54 &1.52 &   $\pm$0.09 &   $\pm$0.16 \\    
 2.335 & 1.49 &  $\pm$0.14  &$\pm$0.16 &1.66 &   $\pm$0.15 &   $\pm$0.18 \\    
\hline
\end{tabular}                                          
\end{center}
\end{table} 
\twocolumn
 

\begin{thebibliography}{1999}

   
\bibitem {Bergdolt} A.M.~Bergdolt et al.,
Phys.~Rev.~{\bf D48} (1993)2969
  
\bibitem {Chiavassa} E.~Chiavassa et al.,
  Phys.~Lett.~ {\bf B322} (1994)270
  
\bibitem {Calen} H.~Cal\'{e}n et al.,
  Phys.~Lett.~ {\bf B366} (1996) 39

  
\bibitem {Hibou} F.~Hibou et al.,
  Phys.~Lett.~ {\bf B438} (1998)41
  
\bibitem {Calen3} H.~Cal\'{e}n et al.,
  Phys.~Lett.~ {\bf B458} (1999)190
  
\bibitem {Smyrski} J.~Smyrski et al.,
  Phys.~Lett.~ {\bf B474} (2000) 182
  
\bibitem {Botovic} M.Batini\'{c}, A. \^{S}varc and T-S. H. Lee,\\ Physica Scripta {\bf 56}
  (1997) 321
  
\bibitem {Bernard} V. Bernard, U. Kaiser, Ulf-G. Mei{\ss}ner,\\
  Eur. Phys. J. {\bf A4} (1999) 259

\bibitem {Faldt} G.F\"aldt and C.Wilkin, Physica Scripta {\bf 64} (2001) 427
  
  
\bibitem {Nakayama} K.~Nakayama,  nucl.th.~0108032(2001) 
  
\bibitem {Pena} M.~T.~Pe\~{n}a, H.~Garcilazo, and D.~Riska,\\ Nucl.Phys. {\bf A683} (2001)
  322

\bibitem {pawelCosy11} P. Moskal et al., Phys. Lett. {\bf B482} (2000) 356 

\bibitem {Target} A.~Hassan et al.,\\  Nucl.~Instr.~and Methods {\bf
    A425}(1999)403
  
\bibitem {Startzaehler}P.~Michel et al.,\\ Nucl.~Instr.~and Methods {\bf
    A408}(1998)453

\bibitem {Quirl} M.~Dahmen et al.,\\ Nucl.~Instr.~and Methods {\bf A348}(1994)97

\bibitem {Barrel} A.~B\"{o}hm et al.,\\
  Nucl.~Instr.~and Methods {\bf A443}(2000)238



\end{thebibliography}
\end{document}